\documentclass[%
 aip,
 amsmath,amssymb,
 reprint,
 twocolumn
]{revtex4-1}

\usepackage[ngerman,english]{babel}
\usepackage{color}
\definecolor{Blue}{rgb}{0.3,0.3,0.9}  
\usepackage{graphicx}
\usepackage{dcolumn}
\usepackage{bm}
\usepackage{hyperref}
\hypersetup{
 colorlinks=true,
    linkcolor=blue,
    citecolor=blue,      
    urlcolor=blue,
}
\usepackage{xcolor}
\usepackage{mdframed}
\usepackage{todonotes}
\graphicspath{{Figures/}}
\usepackage{comment}

\newcommand{\COMMENT}[1]{}
\begin{document}

\title{Impact of Electron Transport Models on Capillary Discharge Plasmas}

\author{A. Diaw }
\email{diaw@radiasoft.net}
\author{ S.\,J. Coleman, N.\,M. Cook, and J. Edelen }
 \affiliation{RadiaSoft LLC, 6640 Gunpark Dr Suite 200
Boulder, CO 80301 U.S.}
\author{E. C. Hansen }
\affiliation{Flash Center for Computational Science, Department of Physics and Astronomy, University of Rochester, 500 Wilson Blvd. PO Box 270171, Rochester, NY 14627 U.S.}
\author{P. Tzeferacos}
 \affiliation{Flash Center for Computational Science, Department of Physics and Astronomy, University of Rochester, 500 Wilson Blvd. PO Box 270171, Rochester, NY 14627 U.S.}
 \affiliation{Laboratory for Laser Energetics, University of Rochester, 250 E. River Rd, Rochester, NY 14623 U.S.
}
\date{\today}

\begin{abstract}
 
Magnetohydrodynamics (MHD) can be used to model capillary discharge waveguides in laser-wakefield accelerators. However, the predictive capability of MHD can suffer due to poor microscopic closure models.
Here, we study the impact of electron heating and thermal conduction on capillary waveguide performance as part of an effort to understand and quantify uncertainties in modeling and designing next-generation plasma accelerators. To do so, we perform two-dimensional high-resolution MHD simulations  using an argon-filled capillary discharge waveguide with three different electron transport coefficients models. The models tested include (i) Davies et al. (ii) Spitzer, and (iii) Epperlein-Haines (EH). We found that the EH model overestimates the electron temperature inside the channel by over $20\%$ while predicting a lower azimuthal magnetic field. Moreover, the Spitzer model, often used in MHD simulations for plasma-based accelerators, predicts a significantly higher electron temperature than the other models suggest.
\end{abstract}

\maketitle

\section{\label{sec:intro} Introduction}

Laser‐wakefield‐accelerators (LWFA) have the technological potential to supplant conventional‐radio‐frequency‐accelerators and also bring about a new generation of compact‐tabletop‐accelerators \cite{Esarey_2009,Kim_2021}. At present, LWFAs can produce stable electron beams with ultrashort duration, GeV-scale energy, and very low emittance from centimeter-scale acceleration stages at high repetition rates \cite{Maier_2020, Salehi_2021, He_2013}. These high energy and brightness electron beams are essential to meeting the demands of future accelerators such as next-generation colliders\cite{Schroeder_2010} or compact free-electron lasers\cite{Maier_2012}. However, improved plasma sources and optics are required to achieve these designs. 

Capillary discharge plasmas offer a promising solution to these challenges, enabling tailored plasma density profiles for a variety of accelerators applications. This approach has been successfully used to generate plasma waveguides to propagate intense laser pulses over many Rayleigh ranges, thereby increasing the achievable peak energy from wakefield acceleration \cite{Leemans_2014,Gonsalves_2019}. Due to their capacity to generate high azimuthal magnetic fields, capillaries have also been employed as compact, efficient lenses for beam transport \cite{vanTilborg_2015,Lindstrom_2018}, energy spread manipulation \cite{DArcy_2019}, and for coupling high brightness electron beams between plasma accelerator stages \cite{Steinke_2016nature}. Accurate plasma density and temperature predictions require capturing electron energy transport and magnetic field propagation within the plasma. Magnetohydrodynamics (MHD) simulations, extended to include non-ideal plasma effects such as gradient-driven transport and electric-current-driven transport, can play a crucial role in modeling these systems \cite{Bobrova_2001, Bagdasarov_2017}.
 
As far as we know, the magnetic field components (perpendicular and cross) impact on the transport phenomena perpendicular to the magnetic field has not been considered in previous capillary plasma simulations. Nevertheless, the presence of a magnetic field reduces the collisional mean-free paths of electrons and ions. As a result, depending on the magnetic field amplitude, plasma transport coefficients in the directions perpendicular to the magnetic field may become very small in such a manner that the associated fluxes are strongly anisotropic. This paper explores the role of anisotropic transport in waveguide applications, specifically the impact of resistive and conductive models on gas-filled capillary discharges. These Coulomb-collision effects impact the time for the capillary to reach steady-state and the final profiles of temperature and magnetic fields that can be achieved.

Various transport coefficient models exist in the literature for physical transport phenomena \cite{1965RvPP....1..205B, 1986PhFl...29.1029E, JiHeld_2013, 2021PhPl...28a2305D, Sadler_2021, Simakov2022}. For weakly coupled plasmas, often these models are obtained by fitting numerical solution of Vlasov-Fokker-Planck (VFP) kinetic equation \cite{Rosenbluth_1957,Larroche_1993} with some functional form of the ionization and magnetization. Braginskii \cite{1965RvPP....1..205B} evaluated the anisotropic coefficients using the first 3-terms in the Laguerre polynomials expansion of the electron distribution function of the Fokker-Planck equation. Braginskii provided fits for a few atomic numbers $Z=1-4$ and various magnetization $\chi=\omega_{ce}\tau$ ($\omega_{ce}$ is the cyclotron frequency and $\tau$ is electron-ion relaxation time) with an accuracy of $20\%$. However, Braginskii fits exhibit substantial inaccuracies, especially at large $\chi$. Later, Epperlein and Haines \cite{1986PhFl...29.1029E} (EH) used the numerical solution of the Fokker-Planck equation and proposed formulas for the transport coefficients for a large number of atomic numbers, thereby fixing the asymptotic behavior displayed by the Braginskii coefficients. However, some of the EH coefficients are discontinuous concerning the atomic number. Ji and Held~\cite{JiHeld_2013} revisited the work of Epperlein and Haines and proposed formulas for arbitrary $Z$. These fits used 160-terms of the Laguerre polynomials expansion and are accurate to within $1.\%$. Recent theoretical and computational work by Davies et al.~\cite{2021PhPl...28a2305D} and Sadler et al.~\cite{Sadler_2021}, extended the work of Epperlein and Haines and Ji and Held \cite{JiHeld_2013}, particularly the resistivity and electrothermal coefficients. Their treatments rely on solving the VFP equation using a large number of terms in the Laguerre expansion \cite{} or on directly numerically solving of the VFP \cite{}. In both works, the authors ensure that the polynomials fits satisfy the asymptotic behavior and that the difference between perpendicular and parallel components of transport models fits are smooth functions in $\chi$ and $Z$. Note that all of these models should recover the Spitzer Lorentz gas results for unmagnetized plasma.

We present here an implementation of various popular electron transport  (Epperlein and Haines \cite{1986PhFl...29.1029E}, and Davies et al. \cite{2021PhPl...28a2305D}) models in the radiation-MHD code FLASH \cite{fryxell2000flash, Tzeferacos:HEDP:2015} and assess their performance on the dynamics argon-filed capillary discharge waveguide. Our simulations show that the EH model strongly underestimates the magnetic field and electron energy. This inhibition continues into the quasi-steady phase. 
Our results demonstrate the sensitivity of the electron energy and magnetic field evolution to collisional phenomena and reinforce the need to benchmark transport models in MHD codes with experiments.

\COMMENT{Various transport models exist in the literature for anisotropic transport. For weakly coupled plasmas, often these models are obtained by fitting numerical solution of Vlasov-Fokker-Planck kinetic equation \cite{Rosenbluth_1957,Larroche_1993} with some functional form of the ionization and magnetization. We present an implementation of various popular electron transport  (Epperlein and Haines \cite{1986PhFl...29.1029E}, and Davies et al. \cite{2021PhPl...28a2305D}) models in the MHD FLASH code. The argon filed capillary discharge waveguide dynamics simulation shows that the EH model strongly underestimates the magnetic field and electron energy. This inhibition continues into the quasi-steady phase. 
Our results demonstrate the sensitivity of the electron energy and magnetic field evolution to collisional phenomena and reinforce the need to benchmark transport models in MHD code with experiments.}

\section{Magnetohydrodynamics Model}

For this study, capillary discharge behavior is examined via fully resolved magneto-hydrodynamics simulations performed using the FLASH code. FLASH\cite{fryxell2000flash} is a publicly available\footnote{For more information on the FLASH code, visit: \url{https://flash.rochester.edu}}, parallel, multi-physics, adaptive-mesh-refinement, finite-volume Eulerian hydrodynamics and MHD\cite{Lee2013} code, whose high energy density physics capabilities\cite{Tzeferacos:HEDP:2015} and synthetic diagnostics \cite{tzeferacos2017numerical} have been validated through benchmarks and code-to-code comparisons~\cite{fatenejad2013collaborative, orban2013radiation}, as well as through direct application to laser-driven laboratory experiments\cite{meinecke2014turbulent, meinecke2015developed, li2016scaled, tzeferacos2018laboratory, chen2020transport, bott2021time, bott2021inefficient, Meinecke2022}.

The evolution of a single-fluid, three-temperature (3T) flow with density, $\rho$, velocity, ${\bf u}$, energy, ${\cal E}$, and magnetic field ${\bf B}$ are given as follows \cite{Tzeferacos:HEDP:2015}:
\COMMENT{The system is governed by an equation of continuity, a momentum equation, energy equations for the fluid and $s$ particles (ion, electrons, radiation) and an induction equation as follows:}
\begin{eqnarray}
   \label{eq:density}
     \frac{\partial  \rho }{ \partial  t} + \nabla \cdot  \big(\rho {\bf u} \big)&=&0, \\
     \label{eq:momentum}
 \frac{\partial \rho {\bf u} }{\partial t}   +\nabla \cdot \big(  \rho {\bf u} {\bf u} 
      + p_t {\bf I} - {\bf B B}\big) &=& 0,\\
      \label{eq:energy}
 \frac{\partial \rho {\cal E}}{\partial t}  +\nabla \cdot \big[\big(\rho {\cal E} + p_t\big) {\bf u} -\big({\bf u}\cdot {\bf B}\big){\bf B} \big]  &\!=\!&   \nabla \cdot \big[\!{\bf B} \!\times \!\big(\eta {\bf j}\!\big)\!\big]
\\ \nonumber && -\nabla \!\cdot \!\boldsymbol{q},\\
 \label{eq:induction}
       \frac{\partial{\bf B}}{\partial t}
    -  \nabla   \times \big({\bf u} \times {\bf B} \big)  + \nabla\times \eta {\bf j} &=& 0,
  \end{eqnarray}
  where the current density is given by ${\bf j}=\nabla \times {\bf B}$. The total specific internal energy is given by ${\cal E}=\rho e_{\rm int}
  +\rho {\bf u}^2/2+ {\bf B}^2/2$, where the internal energy 
  ${\cal E} = e_e+e_i+e_r$ includes the contributions of electron ($e_e$), ions ($e_i$) and radiation ($e_r$). Other variables include the total pressure, $p_t =p_e+p_i+p_r+B^2/2$, which is a sum over electron, ions, radiation and magnetic pressures. Finally, $\eta$ is the resistivity, and the total heat flux, defined as $\boldsymbol{q} =\boldsymbol{q_e}+\boldsymbol{q}_r$ takes both electron thermal conduction ($\boldsymbol{q}_e =\kappa \nabla T_e$) and radiation flux.
  
 The strict local thermodynamic equilibrium assumption is often not justified for all types of laboratory capillary discharges, particularly for low-density systems and during early times. We treat the 3T-components by separately solving for the internal energies of the fluid's electron, ion, and radiation components given by:
 \begin{eqnarray}
      \label{eq:3T:i}
    \frac{\partial \rho e_i }{\partial t} +\nabla \cdot \big(\rho 
    e_i {\bf u}\big) + p_i\nabla\cdot {\bf u}  &=& \rho \nu_{ei}( T_{\rm e}-T_{\rm i}),\\
     \label{eq:3T:e}
        \frac{\partial \rho e_e }{\partial t} +\nabla \cdot \big(\rho 
    e_e {\bf u}\big) + p_e\nabla\cdot {\bf u}  &=& \rho \nu_{ei}( T_{\rm i}-T_{\rm e})\\ \nonumber && -\nabla \cdot \boldsymbol{q}_e-{\cal Q}_r  + {Q}_{\rm Ohm},\\
     \label{eq:3T:r}
            \frac{\partial \rho e_r }{\partial t} +\nabla \cdot \big(\rho 
    e_r {\bf u}\big) + p_r\nabla\cdot {\bf u}  &=& -\nabla \cdot \boldsymbol{q}_r+{\cal Q}_r,
  \end{eqnarray}
 where $\nu_{ei}$ is the electron-ion collisions frequency, ${\cal Q}_r={Q}_{\rm ems}-{Q}_{\rm abs} $ is the total radiation flux, and ${Q}_{\rm Ohm}$ is the internal energy density due to Ohmic heating. The radiation field is followed using a multi-group diffusion approximation. Our simulations employ IONMIX opacity and tabular equation of state data\cite{1989CoPhC..56..259M}. Because we only consider a 2D cylindrical geometry for these comparisons, two of the source terms (Biermann and electrothermal gradients) do not need to be included.

\COMMENT{where the coefficients for Hall ($\delta_*$) and Nernst ($\gamma_*$) corrections are given by:
 \begin{eqnarray}
  \delta_{\perp} &=& \frac{\eta_{\wedge}}{\chi}, \delta_{\wedge}  = \frac{\eta_{\perp}-\eta_{\parallel}}{\chi},\\
   \gamma_{\perp} &=&\frac{\beta_{\wedge}}{\chi} \tau_{ei}, \gamma_{\wedge}  = \frac{\beta_{\perp}-\beta_{\parallel}}{\chi}\tau_{ei},\\
      \hat{\alpha} &=&\frac{\kappa}{\chi}. 
 \end{eqnarray}
 Here $\chi=\omega_{ce}\tau_{ei}$ is the electron magnetization, $\omega_{ce}=eB/m_e$ is the electron gyrofrequency, and $\tau_{ei}$ is the electron-ion relaxation time. Note that the coefficients are function of the magnetization and ionization. The induction equation (\ref{eq:induction}) is composed of an advection term, a diffusion term, a resistivity gradient term.
 The Hall coefficients introduce a correction to the advection velocity. The cross-gradient-Nernst term $\gamma_{\wedge}$  moves field perpendicular to magnetic field and temperature, while the parallel Nernst term  $\gamma_{\parallel}$ moves magnetic field along the temperature gradient.
 It is clear from (\ref{eq:induction}) that the anisotropic terms $\perp$ and $\wedge$ terms act to inhibit advection of the magnetic field.}

FLASH has several logical switches to turn on or off various extended MHD effects, including anisotropic thermal conduction, magnetic resistivity, the Hall term, and the Biermann battery. The code computes three thermal conductivity coefficients $\boldsymbol{\kappa}= [\kappa_{\parallel}, \kappa_{\perp}, \kappa_{\wedge}]^{T}$, which depend on the transport coefficient implementation. The $\kappa_\perp$ coefficient depends on the magnitude of the magnetic field and effectively inhibits the diffusion of heat perpendicular to the field. The $\kappa_\wedge$ coefficient corresponds to the Righi-Leduc effect and has not been determined to be essential for plasma conditions considered in gas-filled capillary discharge waveguide. The thermal diffusion solver in FLASH incorporates these coefficients and solves for the resulting temperature iteratively using the HYPRE library \cite{hypre}. This method is fully-implicit with no time-step restriction. FLASH also computes as many as three magnetic resistivity coefficients $\boldsymbol{\eta}= [\eta_{\parallel}, \eta_{\perp}, \eta_{\wedge}]^{T}$, which depend on the choice of transport coefficient implementation. Magnetic resistivity is essential to computing Ohmic heating, the main driving term in the capillary discharge dynamics. The temperature gradient is perpendicular to the magnetic field for capillary discharges, and the heating term can be written as $Q = \eta_\perp J^2$, where $J$ is the current density. This term is calculated separately like a source term in FLASH and added appropriately to the internal electron energy. Unlike thermal diffusion, FLASH solves for magnetic diffusion with an explicit flux-based method; thus, a time-step restriction is required. A diffusive time of $(\Delta x)^2 / \eta_\perp$ is calculated, where $\Delta x$ is the computational cell width. If the diffusive time is less than the hydrodynamic time, then the diffusive time sets the overall time step of the simulation.

 The induction (\ref{eq:induction}) and fluid energy (\ref{eq:energy}-\ref{eq:3T:r}) equations contain magnetized transport coefficients for resistivity $\boldsymbol{\eta}= [\eta_{\parallel}, \eta_{\perp}, \eta_{\wedge}]^{T}$ and thermal conductivity $\boldsymbol{\kappa}= [\kappa_{\parallel}, \kappa_{\perp}, \kappa_{\wedge}]^{T}$. These transport coefficients can induce self-generated magnetic fields. These fields in turn can significantly change the evolution of the temperature profiles in high-energy density plasmas thereby showing the importance of accurate transport coefficients models. The modeling of transport coefficients is a subject of continued research in plasma physics. In the next section, we will discuss the transport coefficient models available in the literature for magneto-hydrodynamics simulations, including recent transport models \cite{2021PhPl...28a2305D}.

\section{Anisotropic Transport Coefficient Models}

Let us briefly describe the general approach to deriving anisotropic transport coefficients.

For an electron-ion plasma, the evolution of the electron distribution function $f_e$, in the limit of small-angle scattering from binary collisions, is given by the VFP equation as:
\begin{eqnarray}
\frac{\partial f_e}{\partial t }+{\bf v}\cdot \nabla {f_e} -\frac{e}{m_e} \bigg({\bf E}+\frac{{\bf v}\times{\bf B}}{c}\bigg) \frac{\partial  f_e }{\partial {\bf v}} = C(f_e),
\label{eq:VFP}
\end{eqnarray}
where ${\bf E}$ and ${\bf B}$ are the electric and magnetic fields, respectively. $C(f_e)=C_{ee}+C_{ei}$ is the Fokker-Planck collisions operator and contains both electron-electron $C_{ee}$ and electron-ion collisions $C_{ei}$. The electron distribution is expanded in Cartesian tensors using spherical harmonic expansion; the first two terms can be written as follows \cite{}: $f_e=f_0+{\bf v} \cdot {\bf f}_1(v)/v$. Then, substituting this relation into (\ref{eq:VFP}) and integrating over all angles ($\theta$ and $\phi$) yields an integro-differential equation in ${\bf f}_1$:
\begin{eqnarray}
\frac{\partial {\bf f}_1}{\partial t} + \nabla {f_0} -\frac{e{\bf E}}{m_e} \frac{\partial  f_0 }{\partial v} - \frac{{\bf eB}\times{\bf f}_1}{m_ec}&=& C(f_1),
\label{eq:Linear0VFP}
\end{eqnarray}
Assuming that the plasma is near equilibrium, $f_0$ is set to a Maxwellian, 
\begin{eqnarray}
      \label{eq:Max}
    f_0 &=& n_e \bigg(\frac{m_e}{2 \pi k_BT_e }\bigg)^{3/2} \exp\bigg[-\frac{m_e v^2}{2k_BT_{e}}\bigg],
\end{eqnarray}
and then Equation (\ref{eq:Linear0VFP}) is solved for ${\bf f}_1$. With the knowledge of the electron distribution, electric and heat currents are calculated as follows:
\begin{eqnarray}
\label{eq:Lin-electric-current}
{\bf j}  &=& -\frac{4\pi e}{3}  \int v^3 {\bf f}_1  dv,\\
\label{eq:Lin-heat-current}
 \boldsymbol{q} &=& \frac{2\pi m_e}{3} \int v^5 
 {\bf f}_1  dv. 
  \end{eqnarray}
Since ${\bf f}_1$ is a linear function of the fluid quantities $\nabla n, \nabla T$ and ${\bf E}$, one can introduce the Braginskii electric fields and heat current as:
\begin{eqnarray}
\label{eq:E}
en{\bf E}  &=& - \nabla p + \frac{{\bf j}\times {\bf B}}{c}+\frac{\boldsymbol{\eta}\cdot {\bf j}}{en},\\
\label{eq:heat-current}
 \boldsymbol{q} &=& -\boldsymbol{\kappa} \cdot \nabla T.
\end{eqnarray}
Here $\boldsymbol{\eta}$ \COMMENT{$\boldsymbol{\beta}$,} and $\boldsymbol{\kappa}$ are the electric resistivity \COMMENT{and thermoelectric} and thermal conductivity tensors. 
Given an ionization state $Z$ and magnetization $\chi$,  Eq.~(\ref{eq:Linear0VFP}) is solved for $f_1$, in the quasi-steady limit ($\partial_t {\bf f}_1 \approx 0$). The result is then used to calculate the electric (\ref{eq:Lin-electric-current}) and heat currents (\ref{eq:Lin-heat-current}). The transport coefficients $\boldsymbol{\eta}$ \COMMENT{$\boldsymbol{\beta}$,} and $\boldsymbol{\kappa}$ are obtained from (\ref{eq:E}) and (\ref{eq:heat-current}).  The principal difficulty in this procedure is solving for ${\bf f}_1$, and various approaches have been used in the literature.

One of the first transport coefficients were proposed by Spitzer for $Z=1,2,4,16, \infty$, who solved for ${\bf f}_1$ assuming a Lorentz gas and neglecting the magnetic field. Subsequent versions of the Spitzer model have been proposed later where electron-electron collisions were added into the model through the Coulomb logarithm. Later, Braginskii \cite{1965RvPP....1..205B} evaluated the anisotropic coefficients using Laguerre polynomials expansion ${\bf f}_1$ and provided fits for the transport coefficients electrical and thermal conductivity along with Hall, Nernst, Ettinghausen, and Righi-Leduc coefficients. The coefficients are expressed in terms of the magnetization for a given atomic number. Braginskii suggests fits for $Z=1-4$.  The accuracy of the Braginskii transport coefficients has been demonstrated for high-energy-density plasmas encountered in inertial confinement fusion ($n \sim 10^{25}$ cm$^{-3}$ and $T\sim 1$ keV) through extensive comparisons with quantum molecular dynamics calculations \cite{PhysRevE.89.043105,PhysRevE.87.023104}.  However, the Braginskii transport coefficients are incorrect for degenerate partially ionized or high-Z plasmas\cite{1984PhFl...27.1273L}.

Epperlein and Haines~\cite{1986PhFl...29.1029E} provided later more accurate transport coefficients by extending Braginskii to a wider range of atomic number $Z$. However, recent works by Davies, Wen, Ji, and Held \cite{2021PhPl...28a2305D} and Sadler et al. \cite{Sadler_2021} show inaccuracies in Epperlein and Haines fits. Sadler et al. showed that the EH model yields an artificial magnetic dissipation and discontinues when simulating a perturbed direct-drive laser ablation front at low magnetization.
 
The following section will briefly discuss and juxtapose the Epperlein and Haines \cite{1986PhFl...29.1029E} model and Davies, Wen, Ji, and Held \cite{2021PhPl...28a2305D} model.

     \begin{figure}[!t]
  \includegraphics[width=1\linewidth]{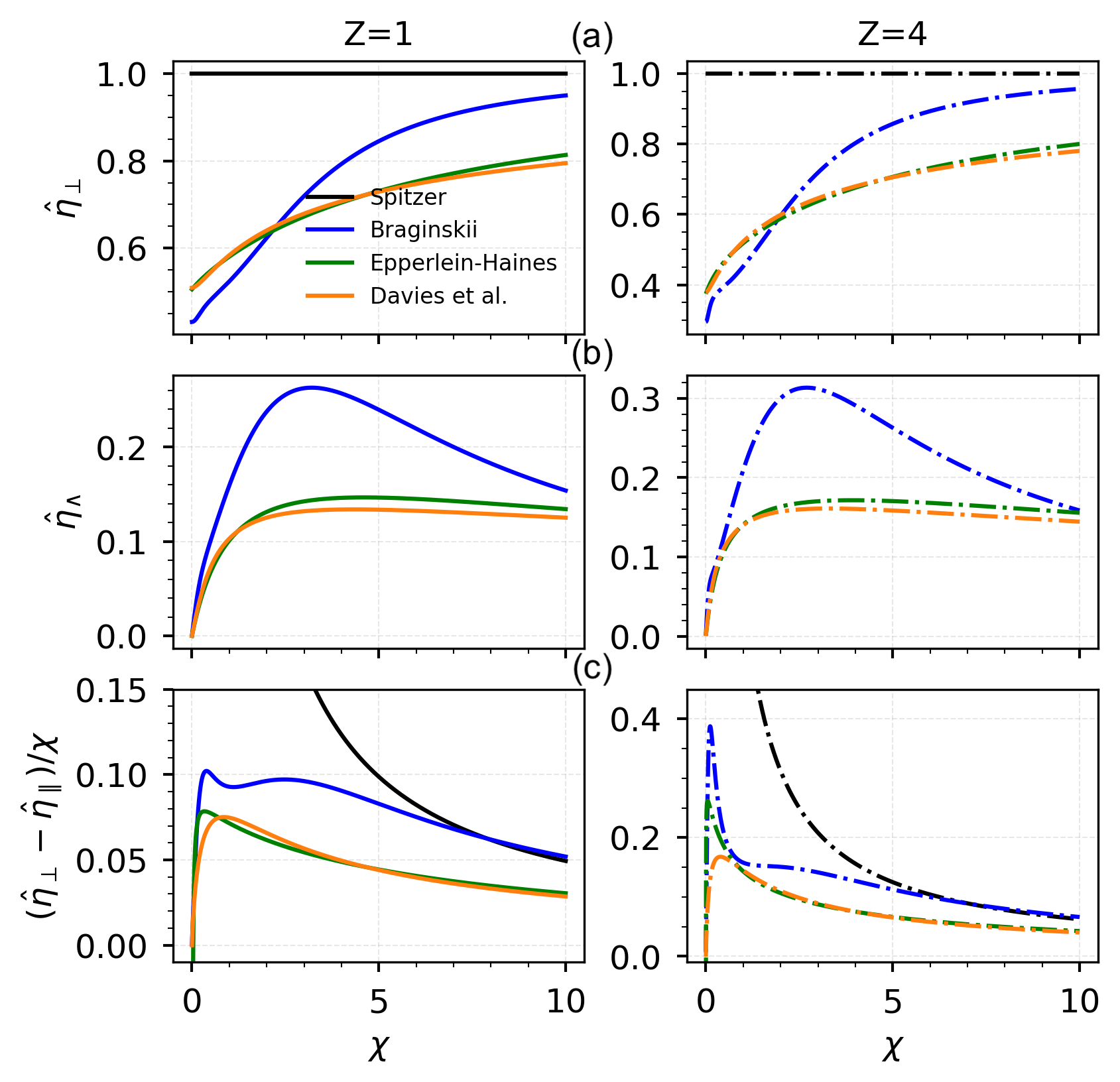}
  \caption{Resistivity coefficients as function of the magnetization from various models for hydrogen and  an argon ($Z=4$) plasma. Solid lines are obtained for $Z=1$, and dashed lines correspond to $Z=4$. These coefficients show the impact of the Ohmic heating and the magnetic field advection. Note although EH and Davies et al. models agree very well for $\hat{\eta}_{\perp}$ and $\hat{\eta}_{\wedge}$,  that their are significant deviation of between them for $\hat{\eta}_{\perp}-\hat{\eta}_{\parallel}$, specially at low $\chi$.}
  \label{fig:resistivity}
  \end{figure}
  
  \begin{figure}[!t]
  \includegraphics[width=1\linewidth]{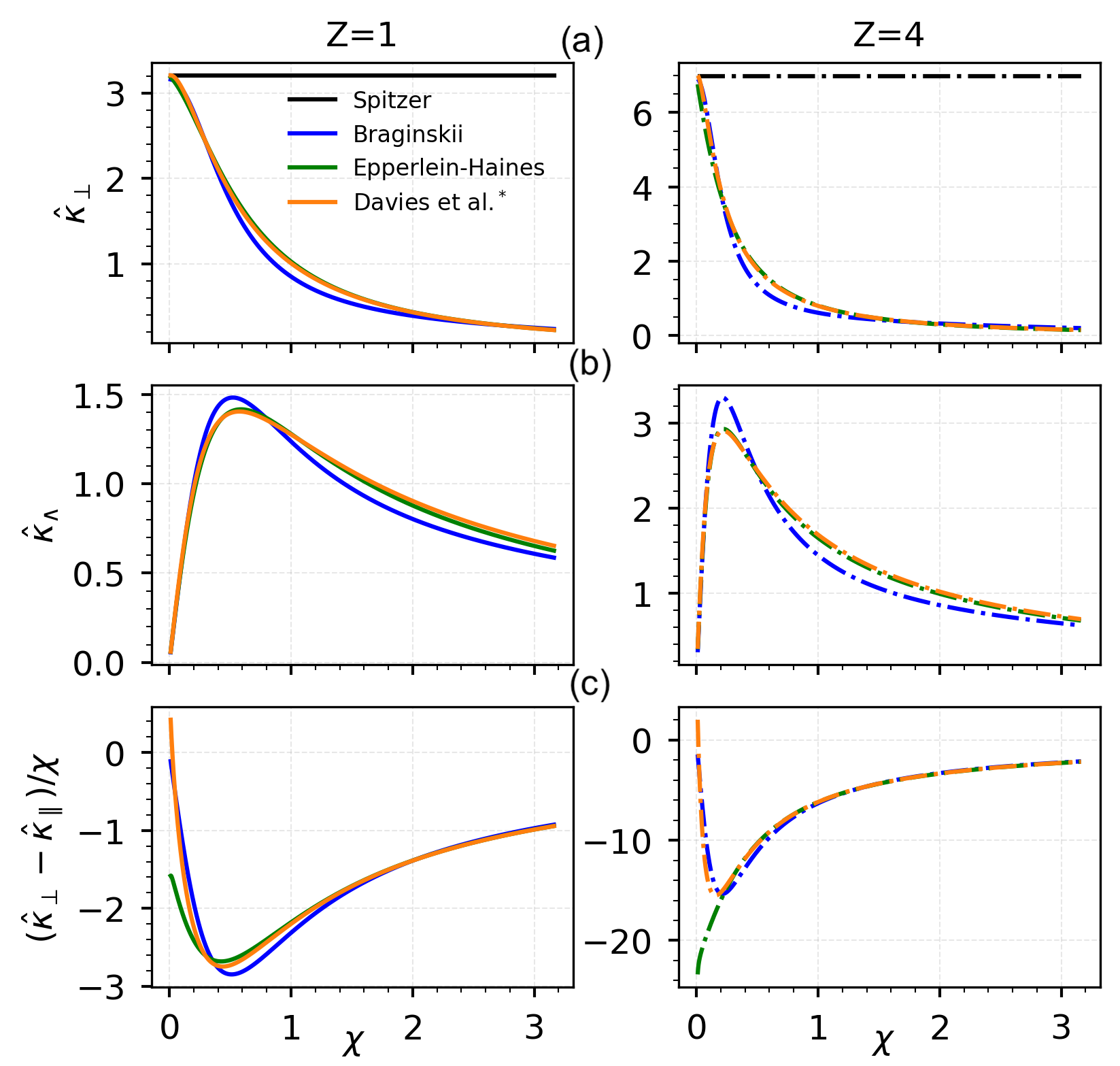}
  \caption{Thermal conductivities coefficients as function of the magnetization from various models for hydrogen and  an argon ($Z=4$) plasma. Solid lines are obtained for $Z=1$, and dashed lines correspond to $Z=4$. We show Spitzer model here is as reference point. Note that all models agree in their predictions for the different components of the conductivities. It is worth noting that we used the Ji and Held~\cite{JiHeld_2013} fits for thermal conductivity coefficients as Davies et al.~\cite{2021PhPl...28a2305D} did not consider thermal conduction.}
  \label{fig:conductivity}
  \end{figure}
  
  \begin{figure*} 
        \includegraphics[width=1.0\linewidth]{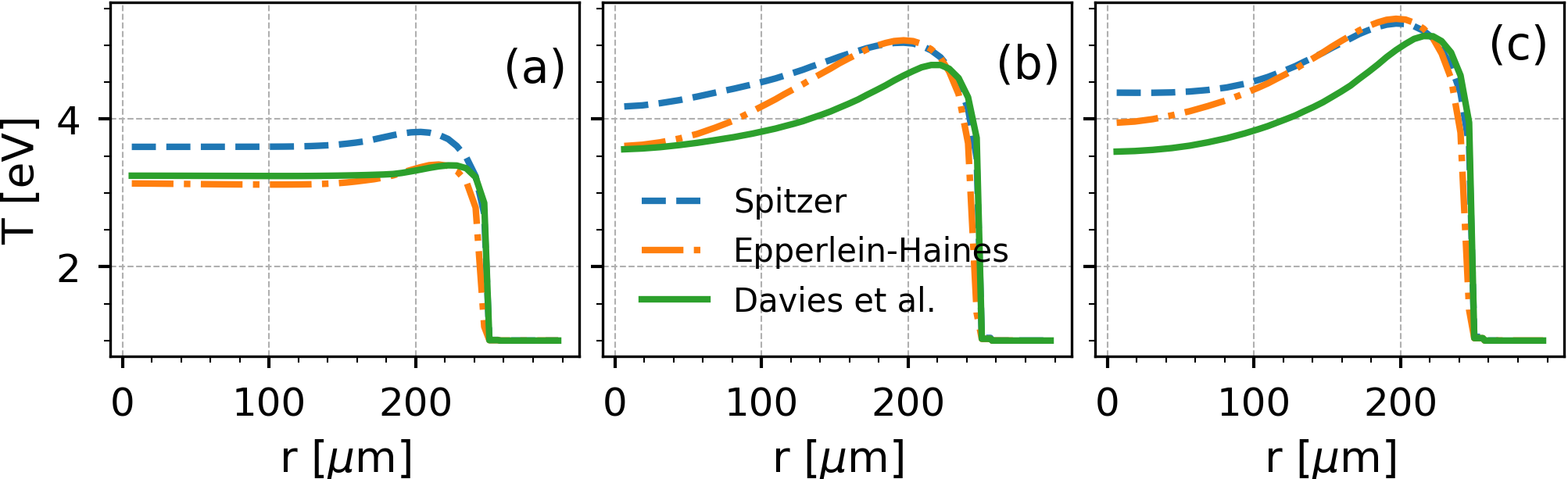}
  \caption{Electron temperatures profiles as a function of the radial coordinate for
   a $400$ microns long capillary-filled Argon with initial density 
$\rho=4.142 \cdot 10^{-5}$ g/cm$^{-3}$ and pressure around $1$ bar. The curves are shown at times $50$, $85$ and $100$ ns.  The solid lines represent the results with the Davies et al. \cite{2021PhPl...28a2305D} transport model, while the dashed lines correspond to the Epperlein and Haines \cite{1986PhFl...29.1029E} transport model. Also shown are results obtained with Spitzer model.}
 \label{fig:temperature}
  \end{figure*}

The first model we considered here is the Epperlein and Haines model, directly derived from the VFP equation. Epperlein and Haines solved Eq.~(\ref{eq:Linear0VFP}) numerically using a finite difference scheme. A five-point differencing scheme was employed to guarantee accuracy. For the velocity grid, they used $100$ points within the interval $[0, v_{max}]$ where $v_{max}=5v_T$ where $v_T$ is the electron thermal velocity $v=(k_BT/m_e)^{1/2}$. The contributions to the distribution function of electrons with velocity $v>v_{max}$ were neglected.
 With this approach, Epperlein and Haines \cite{1986PhFl...29.1029E} provided transport coefficients fits for various atomic numbers in the range of $Z=1-1000$.  Their results showed significant errors with Braginskii coefficients \COMMENT{$\beta_{\wedge}$,} $\kappa_{\perp}$ and $\kappa_{\wedge}$, up to $68\%$. They also demonstrated that $\eta_{\wedge}$ \COMMENT{and $\beta_{\perp}$} vary as $\tau/\chi^{3/2}$ \COMMENT{and $\tau/\chi^{5/3}$} for $\chi \rightarrow \infty$ as opposed to the predictions by Braginskii model of $\tau/\chi$. Here $\tau$ is the electron relaxation time.

While there are a lot of computational benefits of the Epperlein and Haines model as they cover a wide range of atomic number and magnetization, recent works have shown large errors in the resistivity. 

Davies, Wen, Ji and Held \cite{JiHeld_2013, 2021PhPl...28a2305D} revisited Epperlein and Haines's work and solved for ${\bf f}_1$ using Laguerre polynomials expansion with $160$-terms. They proposed formulas for Hall, Nernst, cross-gradient Nernst, and thermoelectric coefficients for any arbitrary $Z$ and $\chi$. Their fits, as well as those of Sadler et al. \cite{PhysRevLett.126.075001}, show significant improvement to the results of Braginskii and Epperlein and Haines. 
The innovation by  Davies et al. and Sadler et al., contrary to Epperlein and Haines, was to make sure the perpendicular resistivity that $\partial_{\chi} \eta_{\perp} (\chi,Z)  = 0 $ as $\chi\rightarrow 0$,

  Let us now show some numerical of comparisons of existing models.
  We present here the change in resistivity and thermal conductivity coefficients with evolving electron magnetization $\chi$. We plot the transport coefficients versus $\chi$, for various $Z$ to estimate the behavior for low-Z and moderate-Z plasmas.  We consider systems consisting of hydrogen and argon gases, respectively. Hydrogen and argon are commonly employed in capillary discharge plasmas for waveguide and lens applications, and their different atomic numbers and ionization states make them valuable candidates for model evaluation. For each quantity, results are compared across models given by Davies et al. (orange), Epperlein and Haines (green), Braginskii (blue), and Spitzer (black). Note that for the electron thermal conductivity transport coefficients, we used the fits by Ji and Held~\cite{JiHeld_2013} as Davies et al.~\cite{2021PhPl...28a2305D} did not consider thermal conduction.
  
  \subsection{Resistivity coefficient}
  
 We first examine the predicted resistivities for these systems as a function of $\chi$, normalized to the values predicted for a classical Lorentz gas model. Fig.~\ref{fig:resistivity} depicts comparative results for hydrogen, $Z=1$, (solid lines) and argon, $Z=4$, (dot-dashed lines) for the four models considered.

The Spitzer model predicts the largest perpendicular resistivity, and sets the upper bound for all models converged to it as $\chi \rightarrow \infty$. EH and Davies et al. give similar predictions for H and Ar and across the considered range of magnetization. The Braginskii formula slightly overestimates the coefficient for $\chi > 1.0$,  but gives lower resistivity (by about $30\%$) for $\chi < 1.0$. 

As for the cross resistivity as shown in Fig.~\ref{fig:resistivity}(b), EH model follows Davies et al. very closely across the parameter space ($\chi$, Z). However, at low magnetization, Braginskii predicts a significantly higher resistivity. It is worth noting that the Spitzer model does not have a cross resistivity component. 

We look now at the difference between the perpendicular and parallel resistivities, $\hat{\eta_{\perp}}-\hat{\eta_{\parallel}}$ in Fig.~\ref{fig:resistivity}(c) as this quantity effectively is used in the induction equation for most extended-MHD formulation \cite{}. When the system is nearly purely unmagnetized $\chi \rightarrow 0$, $(\hat{\eta_{\perp}}-\hat{\eta_{\parallel}})/
\chi$ for all models should approach the Spitzer limit. We can clearly see this behavior in Fig.~\ref{fig:resistivity}(c).  

However, we observe strong disagreement between models at low magnetization $\chi < 1$, most notably for the Ar case. For instance, for $\chi=10^{-1}$ and $Z=4$, Davies et al. predicts a value of $(\hat{\eta_{\perp}}-\hat{\eta_{\parallel}})/\chi \sim 0.18$
while EH and Braginskii suggest a value two times higher. Finally, EH also appears to discontinuously evolve with $\chi$ at low Z for the hydrogen system. This point has been discussed by Sadler et al. \cite{Sadler_2021}.

  \subsection{Thermal conductivity coefficient}
  
 Figure \ref{fig:conductivity}(a) shows the dependence of the non-dimensional thermal conductivities on magnetization $\chi$ for hydrogen, $Z=1$, (solid lines) and argon, $Z=4$, (dot-dashed lines), for the same four models. For $\hat{\kappa}_{\perp}, \hat{\kappa}_{\wedge}$, the various models, except Spitzer, are in relatively good agreement. For $\hat{\kappa}_{\parallel}$ (not shown here), all models agree to the Spitzer model. This result is not surprising, as one constraint used to build these analytical fits is to recover the non-magnetized Spitzer model as $\chi\rightarrow 0$. For the cross-component of the thermal conductivity, Braginskii predicts slightly higher peak amplitude for H while EH and Davies et al. follow each other very closely as depicted in Fig.~\ref{fig:conductivity}(b). Note that this deviation appears to expand for high-Z elements. Finally, for the difference between the quantity $\hat{\eta_{\perp}}-\hat{\eta_{\parallel}}$, we observe a discontinuous evolution from EH prediction for very low magnetization $\chi$ as shown in Fig.~\ref{fig:conductivity}(c).

In summary, there are persistent disagreements between EH, Braginskii and Davies et al. models for the resistivity coefficients, particularly at low magnetization. These differences are large enough to significantly impact Ohm heating and may influence the dynamics in capillary waveguides. For the thermal conductivity coefficients, the Davies et al. model gives similar predictions to Epperlein-Haines's prescriptions, except for $\hat{\eta_{\perp}}-\hat{\eta_{\parallel}}$ where EH presents a discontinuity at lower $\chi$. A detailed discussion about the these transport models can be found in Refs~\cite{PhysRevLett.126.075001,JiHeld_2013, 2021PhPl...28a2305D}.
  
In the next section, we examine the impact of these models on the dynamics of a capillary discharge plasma to identify differences in predictive capabilities for real-world applications.
  
  \begin{figure*}
  \includegraphics[width=1.0\linewidth]{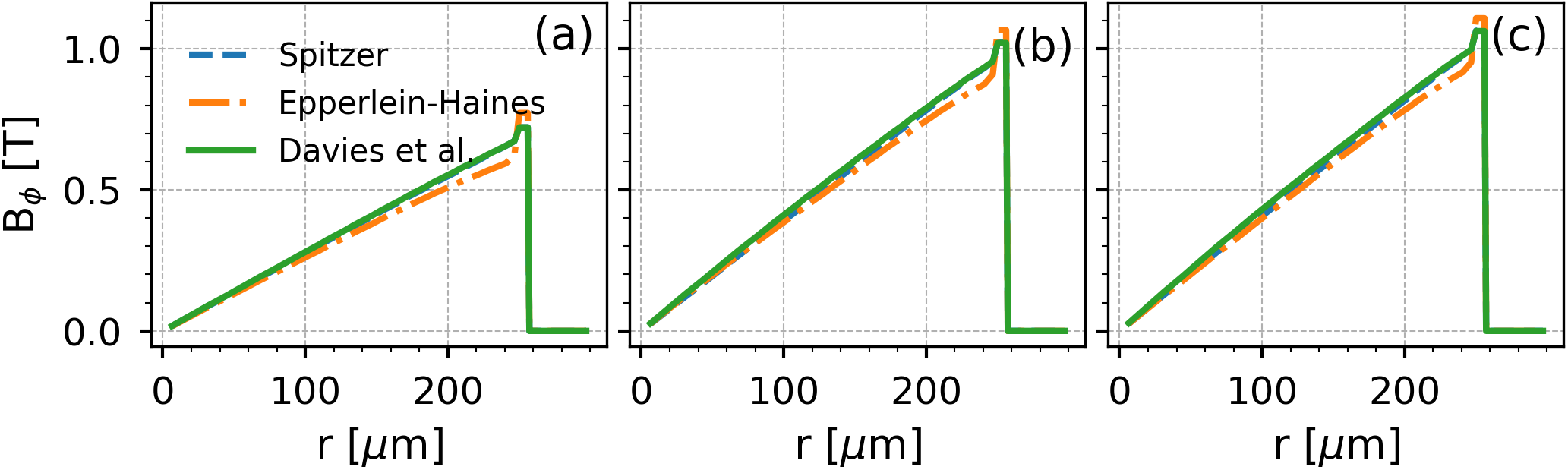}
  \caption{Azimuthal magnetic field profile as a function of the r-coordinate
   for the same parameters as in Fig~\ref{fig:temperature}. Notice that the Epperlein and Haines transport model continuously underestimates the magnetic field within the channel as well as at the wall. Also shown are results obtained with Spitzer model}
 \label{fig:magneticfield}
  \end{figure*}
  
\section{\label{sec:simulations} Gas-filled capillary discharge dynamics}

We now consider the application of these transport models in capturing the dynamics of capillary discharge plasmas, for which magnetic field and temperature evolution are essential. Typical capillaries achieve magnetizations of  $0.05 < \chi < 0.2$, for which these models predict very different resistivities. We consider a capillary consisting of an alumina (Al$_2$O$_3$) cylinder of radius $R$ is filled with an Ar gas of initial density $\rho=4.142 \cdot 10^{-5}$ g/cm$^{-3}$ and initial pressure around $992.6$ mbar; these conditions are comparable to parameters explored experimentally for use with active plasma lenses at accelerator facilities~\cite{Lindstrom_2018}. 

We first describe the capillary discharge model we have developed within the FLASH code~\cite{Cook_2020}. We leverage the cylindrical symmetry of the capillary to perform the simulations in a 2D cylindrical geometry (r-z), although the same configuration can be employed in a 3D Cartesian geometry. Because FLASH is a fluid code, the wall material must be represented as a fluid; however, it is constrained to be immovable while still facilitating thermal and magnetic propagation consistent with the chosen transport model. This approximation permits the inclusion of varying wall geometries within the simulation while retaining boundary dynamics between the plasma flow and the wall.

\begin{figure}[!t]
  \includegraphics[width=0.8\linewidth]{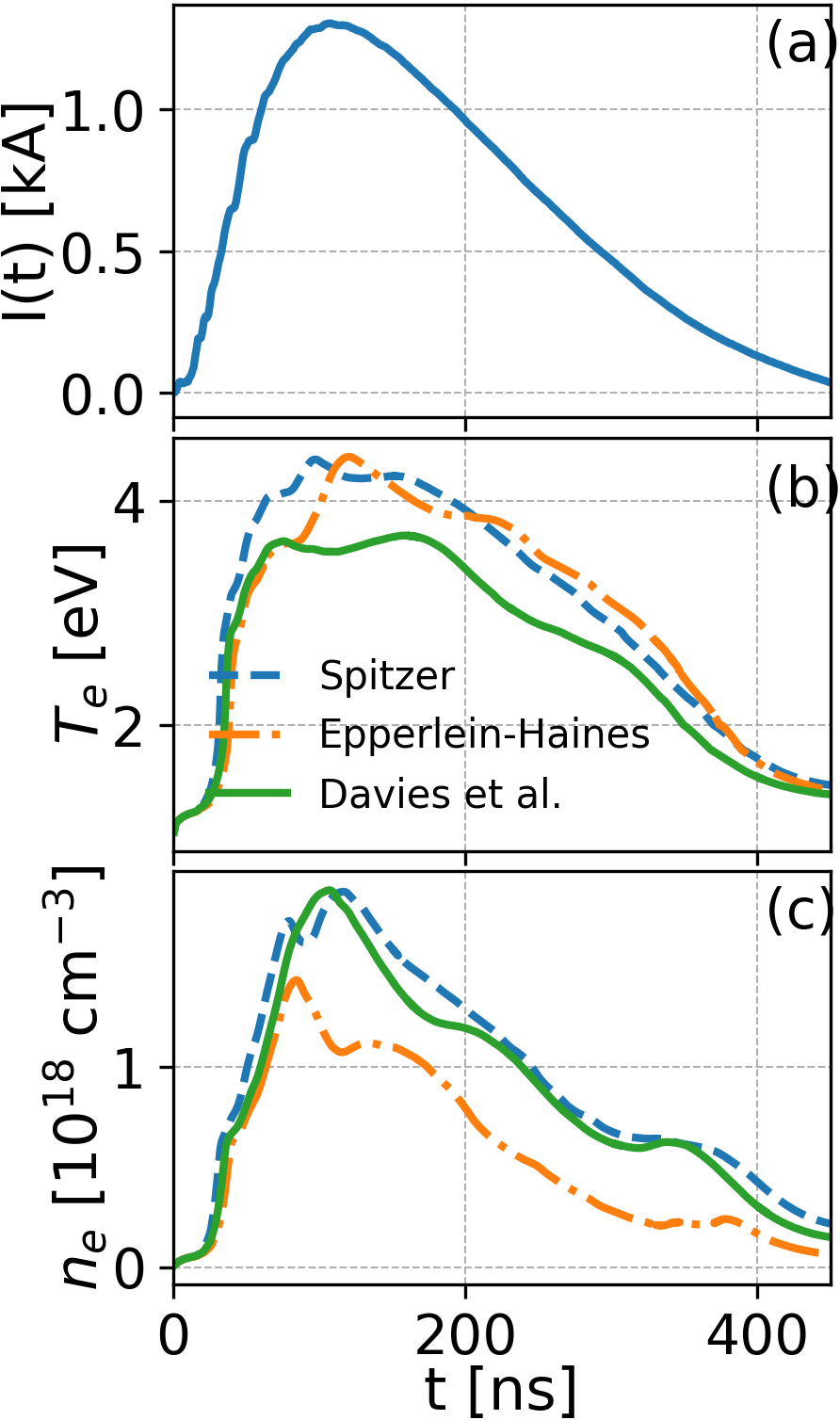}
  \caption{Temporal evolution of (a) the discharge electric current (b) electron temperature (c) electron density for 2D cylindrical geometry FLASH simulations of the argon-filled capillary waveguide. We presented simulated results obtained with various models: Davies et al. (solid lines), Epperlein and Haines (dot-dashed lines), and Spitzer (dashed lines).}
 \label{fig:time_evol}
  \end{figure}

The discharge current is represented through the application of an externally applied magnetic field $B_{\phi}$. Because the capillary wall is insulating, the entire discharge current travels through the inner plasma region of the capillary, generating an azimuthal magnetic field $B_{\phi}$ that satisfies Ampere's Law at $r=R$. This enables exact specification of the magnetic field at the wall interface. The magnetic field is computed as a boundary condition at this interface, and permitted to permeate into the plasma region self-consistently. FLASH's MHD solver explicitly computes the magnetic field on an auxiliary grid using a second-order integrator for each timestep. For the simulations discussed below, the current discharge time-dependence mimics an experimental profile, and reaches a peak value of $1.3$ kA at around $100$ ns, as shown in Fig.~\ref{fig:time_evol}.

We performed three different sets of simulations using Spitzer, Epperlein-Haines, and Davies et al. to compute transport coefficients. Initial temperatures for the electron, ion, and radiation populations are set to $T=1.0$ eV. The averaged ionization and equation of state were estimated using a table generated by the IONMIX code.
  
Figure \ref{fig:temperature} shows the simulated electron temperature profile 
   inside the capillary at times 
   $t=50$, $85$ and $100$ ns. The solid lines represent the results with the Davies et al.~\cite{2021PhPl...28a2305D} model. The predicted electron temperature within the channel by the EH \cite{1986PhFl...29.1029E} model is also shown in dashed lines. We have also displayed the results with Spitzer~\cite{Spitzer}
   isotropic transport model for reference. Since this model has been used in several simulation studies of capillary waveguides~\cite{Bobrova_2001,vanTilborg_2015,Bagdasarov_2021}, it provides a meaningful comparison against the anisotropic models.
      We observed that the electron temperature approximately follows the same trajectory for the three transport models. However, one notices that the EH model overestimates the temperature within the channel by over $20\%$. Our results also show that the Spitzer model consistently predicts a hotter channel. This can be explained by the fact that the Spitzer model does not account for the magnetic field, which acts to slow electrons motion, resulting in more resistive heating. Note that the magnetization $\chi$ value will rise due to plasma heating and the magnetic field transport. 

The azimuthal magnetic field is plotted in Fig.~\ref{fig:magneticfield}. Simulated fluid results using Spitzer, EH and Davies et al. transport models are shown at times three different times. It clearly shows good agreement between the Spitzer and Davies et al. new transport fits. EH predicts a slightly reduced magnetic field across the channel, and the discrepancy appears to increase near the wall. 

The temporal evolution of the temperature and electron density are depicted in Fig.~\ref{fig:time_evol} for the three different transport models. At time $t \sim 100$ ns, corresponding to current peak, Spitzer and EH predicts a higher electron temperature but lower density than the modern transport coefficients from Davies et al. Finally, we note that the Ar gas is only partially ionized, with a maximum average ionization around $\bar{Z} \sim 2.39$ at $T\sim 4.0$ eV.

\section{Conclusions}\label{sec:conclusions}

We have investigated the influence of fully anisotropic resistive heating and thermal conduction transport models on the evolution of gas-filled capillaries discharge plasmas, emphasizing applications in laser-plasma accelerators and electron beam transport. We considered three separate models, Spitzer, Epperlein \& Haines, and Davies et al., and implemented them each within self-consistent simulations of capillary discharge plasmas using the FLASH code. Significant variations were observed in the predicted resistivities across a range of typical operating conditions, resulting in substantive differences in Joule heating and thermal transport throughout the capillary. 

These discrepancies may result in significant deviations in critical figures of merit for these devices. For an argon-filled capillary, the Epperlein \& Haines model overestimates the temperature and underestimates the peak density by more than $20\%$ compared to the transport model recently proposed by Davies et al., while predicting a reduced azimuthal magnetic field within the channel. Likewise, the isotropic Spitzer model used in most MHD studies for capillary discharges shows a similar overestimation of the electron temperature within the capillary compared to modern transport coefficients from Davies et al. These results have considerable ramifications on the choice of tunable operating parameters such as the background gas density and discharge current.

\section*{Acknowledgments}

This work was supported by the U.S. Department of Energy (DOE), Office of Science, Office of High Energy Physics under Award Number DE-SC0018719. The Flash Center for Computational Science acknowledges support by the U.S. DOE National Nuclear Security Administration (NNSA) under Subcontracts No. 536203 and 630138 with Los Alamos National Laboratory, Subcontract B632670 with LLNL, and support from the Cooperative Agreement DE-NA0003856 to the Laboratory for Laser Energetics University of Rochester. Support from the U.S. DOE ARPA-E under Award DE-AR0001272 is also acknowledged. The software used in this work was developed in part by the U.S. DOE NNSA- and U.S. DOE Office of Science- supported Flash Center for Computational Science at the University of Chicago and the University of Rochester. We would also like to acknowledge helpful discussions and input on discharge current profiles from Dr. Gregory Boyle.

\appendix

\bibliography{paper}

\appendix 
\end{document}